*Original Article*

# Chauhan Weighted Trajectory Analysis reduces sample size requirements and expedites time-to-efficacy signals in advanced cancer clinical trials

**Utkarsh Chauhan [1], Daylen Mackey [2], and John R. Mackey [1,*]**

[1] Faculty of Medicine and Dentistry, University of Alberta, Edmonton, AB, T6G 2R7, Canada; uchauhan@ualberta.ca

[2] CW Trial Analytics, Edmonton, Alberta, Canada; daylen.j.mackey@gmail.com

* Correspondence: jmackey@ualberta.ca

**Abstract:** 1) Background: As Kaplan-Meier (KM) analysis is limited to single unidirectional endpoints, most advanced cancer randomize d clinical trials (RCTs) are powered for either progression free survival (PFS) or overall survival (OS). This discards efficacy information carried by partial responses, complete responses, and stable disease that frequently precede progressive disease and death. Chauhan Weighted Trajectory Analysis (CWTA) is a generalization of KM that simultaneously assesses multiple rank-ordered endpoints. We hypothesized that CWTA could use this efficacy information to reduce sample size requirements and expedite efficacy signals in advanced cancer trials. 2) Methods: We performed 100-fold and 1000-fold simulations of solid tumour systemic therapy RCTs with health statuses rank ordered from complete response (Stage 0) to death (Stage 4). At increments of sample size and hazard ratio, we compared KM PFS and OS with CWTA for (i) sample size requirements to achieve a power of 0.8 and (ii) time-to-first significant efficacy signal. 3) Results: CWTA consistently demonstrated greater power, and reduced sample size requirements by 18% to 35% compared to KM PFS and 14% to 20% compared to KM OS. CWTA also expedited time-to-efficacy signals 2- to 6-fold. 4) Conclusion: CWTA, by incorporating all efficacy signals in the cancer treatment trajectory, provides clinically relevant reduction in required sample size and meaningfully expedites the efficacy signals of cancer treatments compared to KM PFS and KM OS. Using CWTA rather than KM as the primary trial outcome has the potential to meaningfully reduce the numbers of patients, trial duration, and costs to evaluate therapies in advanced cancer.

## 1. Introduction

The Kaplan-Meier analysis with logrank testing is the gold-standard method for analysis of efficacy outcomes in cancer trials. The KM estimator [1] has many advantages: it is a nonparametric method, making it suitable for analyzing time-to-event data that may not follow a normal distribution, and it can handle censored data effectively, where some individuals may not experience the event of interest by the end of the study, ensuring that these individuals are appropriately accounted for in the analysis. The KM estimator is typically used in conjunction with the logrank test [2], which allows hypothesis testing to compare survival curves between different treatment groups in clinical trials. While the KM estimator provides a robust and widely accepted method for estimating survival functions from lifetime data, it does have several important limitations that reduce its utility. In particular, KM analyzes only a single time-dependent endpoint for each trial participant and cannot provide a single analysis that incorporates the many key efficacy endpoints that may occur during a cancer patient's clinical course, including disease response, progression, and death. Furthermore, KM analysis can only model unidirectional outcomes (for example, progression of disease) and cannot model the bi-directional outcomes such as initial disease response, followed by later disease progression, and eventual death, that represent the typical illness trajectory of patients with advanced cancer on clinical trials.

The limitation of the current standard of KM dependent trial designs is reflected in the extensive advanced cancer published literature, in which various individual clinical trial endpoints are assessed for their relative contributions to the actual clinical trajectory of the patient, and their relative utilities as the single endpoint for studying new therapies [3-7]. This literature highlights a clear methodologic gap in existing trial design and analysis techniques.

Chauhan Weighted Trajectory Analysis (CWTA) [8] was specifically developed to address the limitations of KM analysis, while retaining an intuitive visual and test statistic output that would be easily interpreted by clinical and regulatory communities due to its similarity to KM analysis. A generalization of the KM methodology, CWTA has several advantages over traditional KM analysis. CWTA permits the assessment of outcomes defined by various ordinal grades or stages or clinical severity, which are common in medical settings but challenging to analyze using traditional methods like the KM estimator. Importantly, it facilitates continued analysis following changes in health state which may be bidirectional (disease recovery or exacerbation), providing a more comprehensive view of the trajectory of clinical outcomes. Additionally, we proved the general case that using CWTA with multiple efficacy endpoints tended to reduce the samples sizes normally required for KM analysis to demonstrate a difference between treatment arms in a clinical trial [8].

CWTA also retains the merits of KM analysis. It is a non-parametric method with the ability to censor patients who withdraw or are lost to follow-up, ensuring robust analysis of clinical outcomes. CWTA, like KM, provides a graphical summary plot that visually depicts the trajectories of patients over time, making it easy to interpret and compare outcomes between different treatment arms. This is particularly important in the setting of clinical oncology, where oncologists have become accustomed to seeing a single plot for efficacy endpoints of clinical trials, in which the separation of the control and experimental curves conveys meaningful information as to the differential effects of treatment through time. Finally, CWTA introduces a weighted logrank test, a modification of the traditional logrank test, to assess the statistical significance of differences in trajectories between groups, providing a rigorous method for hypothesis testing that is also a familiar analog for clinical oncologists.

Given these theoretical and practical advantages, we hypothesized that CWTA could productively be applied to the setting of RCTs in advanced cancer and would reduce sample size requirements and expedite the efficacy signals when compared to the standard KM analysis with a single efficacy outcome per patient.

## 2. Methods

*2.1 Definitions*

Progression free survival (PFS) is an endpoint that is standardly defined in advanced cancer clinical trials as the time from random assignment in a clinical trial to disease progression or death from any cause. Overall survival (OS) is defined as the time from random assignment to death due to any cause. Partial response (PR), complete response (CR), stable disease (SD), and progressive disease (PD) were defined as per RECIST 1.1 criteria [9].

*2.2 Simulation Design*

The simulation studies were generated using Python 3.8 [10]. Study simulations were stochastic processes in which randomly generated numbers are programmed to mirror fluctuating solid tumour response to systemic therapy cycles with monthly measurements of treatment efficacy. We performed 100-fold and 1000-fold simulations of RCTs in advanced cancer. RCTs were stochastically generated at defined sample size (SS) allocated 1:1 to control or experimental and run for 60 months. Rank ordered health statuses were CR = 0, PR = 1, SD = 2, PD = 3, and Death = 4 (see **Table 1**). All patients we assigned SD at time 0 and, each month, were capable of a single level transition as specified in Table 1: either response (PR then CR), maintained SD, or irreversible exacerbation to PD or Death (see **Figure 1** for example trajectories). Event probabilities were modified between groups as defined by a hazard ratio (HR). Through iterative programming of baseline transition probabilities, we modeled a control group CR rate of ~5% and a PR rate of ~30% to reflect first-line advanced cancer RCTs. We modeled a dropout rate of 10% uniformly distributed over 60 months.

**Table 1.** Ordinal variables used for advanced cancer clinical trial outcomes.

| Health Status | Ordinal Value | Type | Antecedent Transitions |
|---|---|---|---|
| Tumor in complete response (CR) | 0 | Reversible | Recovery after PR (1→0) |
| Tumor in partial response (PR) | 1 | Reversible | Recovery from SD (2→1) Exacerbation following CR (0→1) |
| Stable disease (at baseline; SD) | 2 | Reversible | Baseline condition Exacerbation following PR (1→2) |
| Progressive disease (PD) | 3 | Non reversible | Exacerbation of SD (2→3) |
| Death (all causes) | 4 | Non reversible / Absorptive | Exacerbation of PD (3→4) |

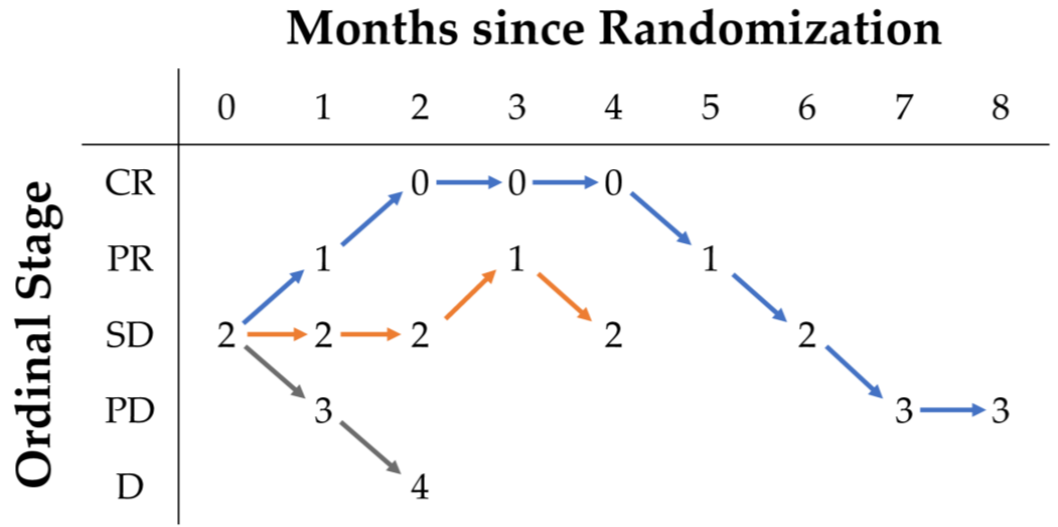

**Figure 1.** Example Disease Trajectories. Patients are enrolled in simulated trials with stable disease (SD) and, through probabilities modified by hazard ratios and time since randomization, may experience partial or complete response to systemic therapy (PR and CR), disease progression (PD) and death (D). Shown are three sample trajectories: in blue, a patient recovers to complete response but gradually worsens to progressive disease; in orange, a patient has stable disease aside from a brief partial response; in dark grey, a patient rapidly experiences disease progression and death.

*2.3 Statistical Analysis*

We performed Kaplan–Meier estimation using the Python 3.8 library "lifelines" [11]. This library was used to conduct logrank tests for KM PFS and OS. Results were validated by assessing the source code for accuracy and making a direct comparison to results from SPSS v26 (IBM Corp., Armonk, NY, USA) [12]. CWTA was performed as described in Chauhan et al. 2023 [3] augmented by a cloud-native cluster, with workloads executed in parallel across optimized Google Compute Engine virtual machines [13].

We ran 1000-fold analyses of completed trials programmed across sample sizes from 20 to 500 and hazard ratios from 0.5 to 0.8. For each trial, a p-value was computed using CWTA, KM PFS and KM OS. The fraction of tests that were significant (at $\alpha < 0.05$) represents the power of the test (correctly rejecting the null hypothesis that the two groups are the same). At each hazard ratio, a sample size requirement for each of the three methods was interpolated as the threshold to reach 0.8 power, a common standard in RCT design.

Time-to-first significant efficacy was determined by calculating the mean and standard deviation of a 100-fold, and for higher HR runs, 1000-fold, simulations programmed across increments of sample sizes from 30 to 500 and hazard ratios from 0.5 to 0.8. Each simulated trial was incremented monthly up to 60 months and analyzed using CWTA (weighted logrank), KM PFS and KM OS (logrank) to tabulate whether, at each monthly time point, statistical significance was reached ($\alpha < 0.05$). The trials that did not reach significance were omitted from the analysis. The analysis pipeline is summarized in **Figure 2**.

To assess how output results varied with input assumptions, sensitivity analyses were also performed with a control group CR rate of ~10% and a PR rate of ~50% to reflect advanced cancer RCTs in which the control group had highly effective therapies.

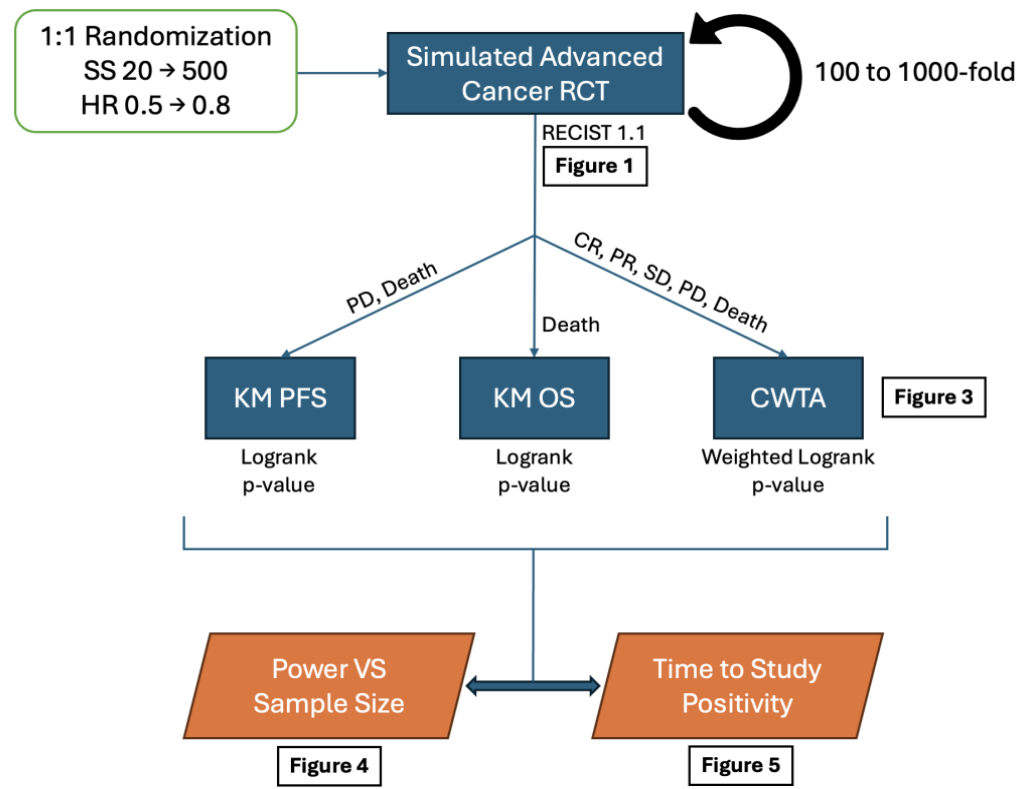

**Figure 2.** A block diagram overview of the study methodology. Advanced cancer randomized controlled trials (RCTs) were stochastically generated at increments of defined sample size (SS) and hazard ratio (HR) allocated 1:1 to the control or experimental group. Health statuses were rank-ordered as per RECIST 1.1 criteria [9]. Each trial was analyzed using Kaplan-Meier Progression-Free Survival (PFS) and Overall Survival (OS) using the logrank test and Chauhan Weighted Trajectory Analysis (CWTA) using the weighted logrank test. 100- and 1000-fold simulations were performed to directly compare the three methods for power (by interpolation of sample size at 0.8 power) and time to study positivity (by incrementing time and assessing first time of $p < 0.05$). The diagram also outlines which Figures of the manuscript correspond to different levels of analysis.

## 3. Results

A single trial analyzed using KM OS, KM PFS, and CWTA is depicted in **Figure 3**. Simulations confirmed CWTA was more powerful than KM analysis and reduced the required sample size to achieve 0.8 power across all hazard ratios (**Table 2, Figure 4).** For a typically sought hazard ratio of 0.8, CWTA reduced sample size by 18% to 35% when compared to KM PFS across a range of hazard ratios, and 14% to 20% when compared to KM OS. The largest sample size reductions came in those trials with higher simulated hazard ratios (that is, less effective interventions). We performed a sensitivity analysis with a higher efficacy control arm (**Table S2, Figure S1**), confirming similar samples size reductions (9% to 22%) and confirming the largest reductions were seen in the higher simulated hazard ratios.

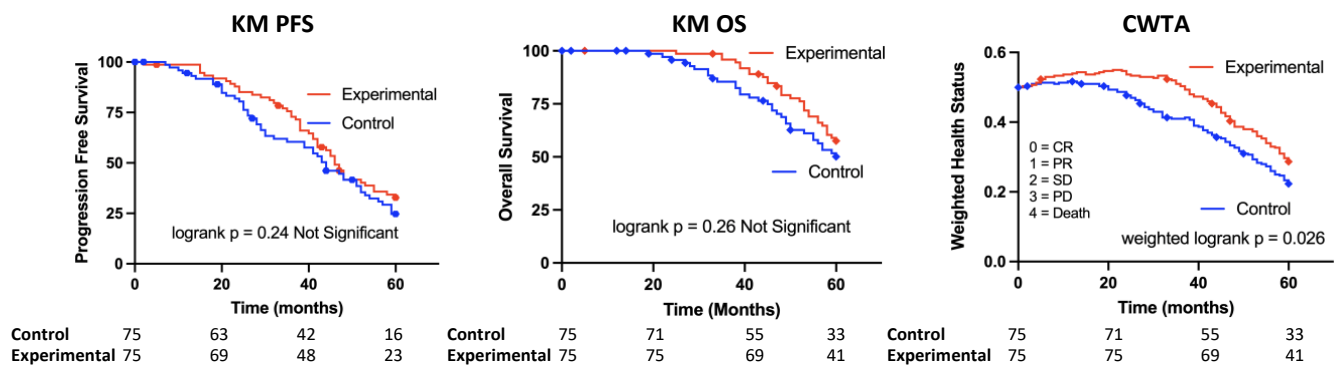

**Figure 3.** An identical advanced cancer trial analyzed three ways: using Kaplan Meier (KM) Overall Survival, KM Progression Free Survival, and Chauhan Weighted Trajectory Analysis (CWTA). Input parameters include a sample size of 150, hazard ratio of 0.7, and a dropout rate of 10% uniformly distributed over a trial duration of 60 months. CWTA models partial and complete responses to therapy, and this added patient information facilitates greater divergence of plotted curves and statistical significance (p = 0.026).

**Table 2.** 80% Power Sample Size Interpolation for Kaplan-Meier estimation of Progression-Free Survival and Overall Survival, and Chauhan Weighted Trajectory Analysis (CWTA) across a range of hazard ratios, using moderately effective control arm assumptions. CWTA reduces the required sample size for 80% power by 18% to 35% versus KM PFS and 14% to 20% versus KM OS.

| HR | Sample Size required for 80% Power | | | Sample Size Reduction using CWTA | |
|---|---|---|---|---|---|
| | CWTA | PFS | OS | vs PFS | vs OS |
| 0.5 | 54 | 66 | 63 | 18% | 14% |
| 0.6 | 69 | 100 | 83 | 31% | 17% |
| 0.7 | 130 | 192 | 162 | 32% | 20% |
| 0.8 | 317 | 486 | 395 | 35% | 20% |

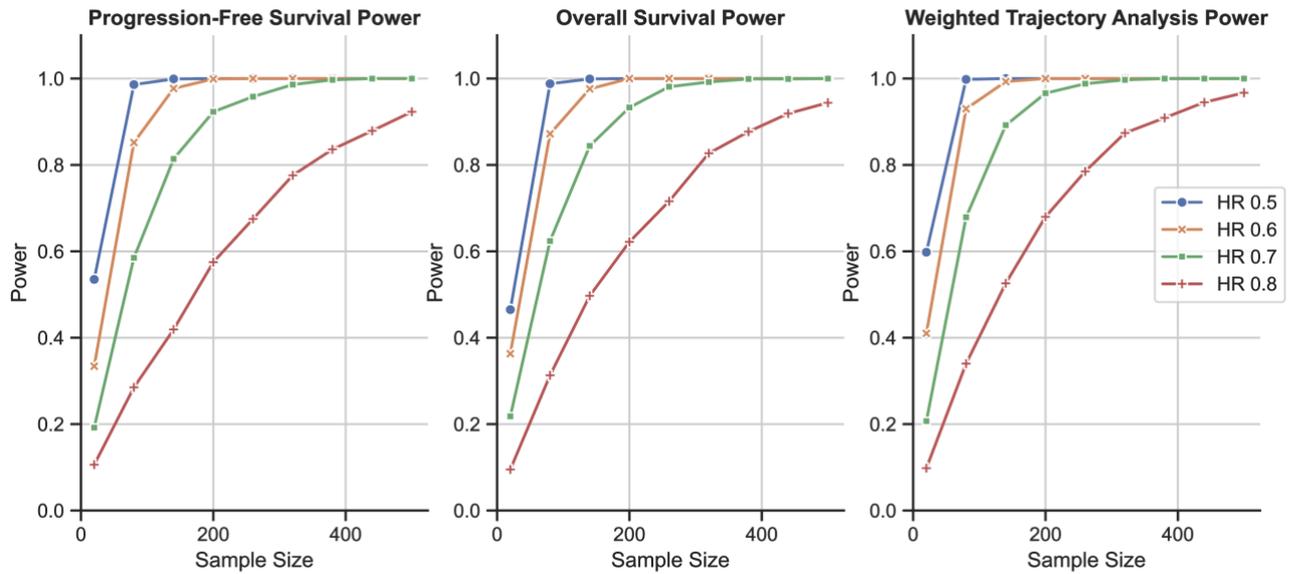

**Figure 4.** Power as a function of sample size for Kaplan-Meier estimation of Progression-Free Survival and Overall Survival, and Chauhan Weighted Trajectory Analysis (CWTA). Using the assumptions of a moderately effective control arm, we performed 1000-fold simulations across a range of hazard ratios. CWTA demonstrated consistently higher power, reflecting a smaller sample size requirement during trial design. Assuming a trial designed to achieve a power of 0.8, CWTA reduced sample size requirements by 18% to 35% when compared to KM PFS, and 14% to 20% when compared to KM OS; the sample size reductions were most marked for studies designed for interventions with higher hazard ratios (that is, less effective interventions).

When trials of a pre-specified sample size were run, CWTA markedly expedited the time-to-efficacy signals when compared to KM PFS and KM OS (**Figure 5, Table S1A**). This effect was seen across a broad range of sample sizes and hazard ratios and confirmed in a sensitivity analysis with a highly active control arm (**Figure S2, Table S1B**). In the extreme cases with a highly effective therapy (low HR intervention), CWTA produced significant RCT outcomes in 20% of the time required to achieve a positive KM OS signal, although more typical was a 50% reduction in time-to-efficacy signal for KM OS and 40% reduction in time-to-efficacy signal for KM PFS. Time-to-efficacy was most reduced in studies with low hazard ratios (that is, highly effective interventions).

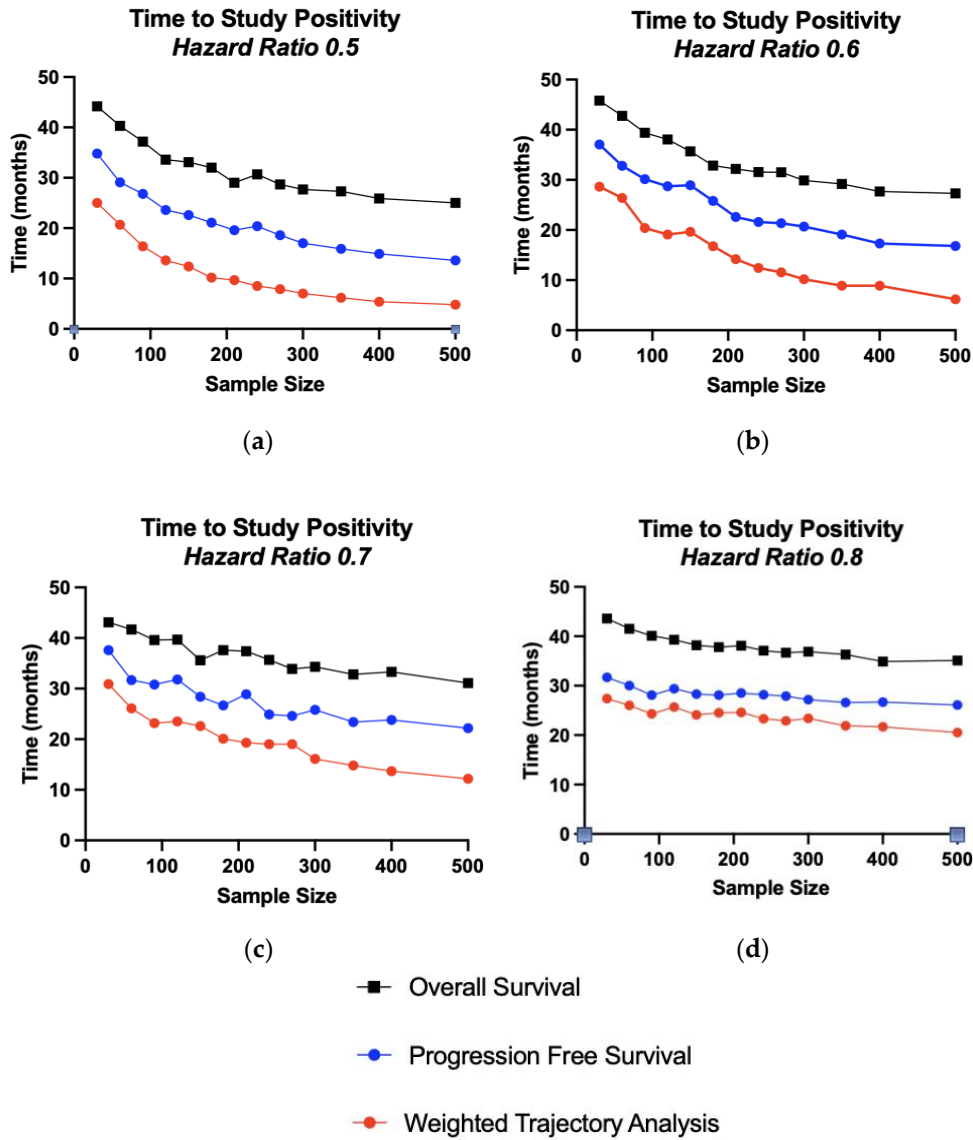

**Figure 5.** Time-to-efficacy signals comparison. We performed one hundredfold / thousandfold simulations of time to study positivity as a function of sample size for Kaplan-Meier estimation of Progression Free Survival and Overall Survival, and for Chauhan Weighted Trajectory Analysis (CWTA) for a range of hazard ratios. CWTA demonstrated consistently shorter time to study positivity; this effect was most profound in studies with a highly effective intervention (that is, a low hazard ratio). (**a**) Time to study positivity at Hazard Ratio 0.5; (**b**) Time to study positivity at Hazard Ratio 0.6; (**c**) Time to study positivity at Hazard Ratio 0.7; (**d**) Time to study positivity at Hazard Ratio 0.8.

## 4. Discussion

In cancer clinical trials, the evaluation of tumor response, progression-free survival (PFS), and overall survival (OS) are interrelated outcome metrics that collectively inform the assessment of treatment efficacy. Solid tumor response, typically measured using the Response Evaluation Criteria in Solid Tumors (RECIST) [9], assesses the change in tumor size and the presence of new lesions as a direct response to systemic therapy. This metric is critical in the early phases (Phase I and IIb) of clinical trials to determine whether a drug has sufficient biological activity to warrant further study, and a strong response to therapy often correlates with longer PFS [14]. PFS is valued in clinical studies as it is considered a surrogate marker for therapeutic benefit, especially in settings where overall survival improvements might be difficult to demonstrate due to crossover designs and subsequent therapies. Studies have shown a correlation between prolonged PFS and increased overall survival [15]. While these relationships generally hold true, the specific relationships amongst response, PFS, and OS may depend on the tumor or tumor subtype studied, the line of therapy, and the availability of effective salvage regimens on progression [16, 17].

In this study, we performed simulations of therapeutic trials conducted in the setting of advanced cancer, and analysed the studies with the traditional KM methodology, as well as its novel generalization, CWTA. By virtue of the ability of CWTA to incorporate and analyze multiple rank-ordered health status endpoints that fluctuate, in this specific case including complete response, partial response, stable disease, disease progression, and death, CWTA outperformed the traditional KM analysis for both PFS and OS. These findings were robust across a representative range of trial sample sizes, and across the range of effect sizes typically sought and observed in advanced cancer trials.

The most dramatic demonstration of improved performance of CWTA over KM is the meaningful reduction in sample size requirements to achieve the typical power threshold of 0.8; sample size reductions of 9% to 35% were noted across the spectrum of study designs (**Figure 4, Table 2, Figure S1, Table S2**). In the setting of advanced cancer, smaller sample requirements are very desirable. Smaller trials are less costly and accrue and complete more rapidly. Increasingly, molecularly targeted subpopulations of cancer patients are being studied, where the drug target may have a prevalence of only a few percentage of the population, and large trials become unfeasible. Similarly, some cancer populations have multiple competing studies open, and sample sizes become critical determinants of a study's chance of success. Proving or disproving efficacy in a smaller population of study participants also spares ineffective therapy and unnecessary followup.

There is a direct relationship between sample size and the cost of an advanced cancer clinical trial. As the sample size increases, the cost of the clinical trial also increases due to direct increases in the costs for each of the following i) participant recruitment outreach efforts, screening processes, and incentives, ii) monitoring and data collection resources for monitoring participant health and ensuring adherence to the trial protocol, iii) logistics and infrastructure for handling the participants including additional clinical sites and staff, and iv) treatment, on-treatment care, and long-term follow-up. A study published by the Tufts Center for the Study of Drug Development demonstrated that each additional patient in a Phase III oncology trial can increase the cost by approximately $36,000 USD [18].

Another demonstration of the improved performance of CWTA over KM is manifest as statistically and clinically significant improvements in time-to-efficacy signals generated by the study, analyzed with both methodologies. The magnitude of these benefits (2- to 6-fold faster time to study positivity; see **Figure 5 and Table S1**) has pragmatic implications for the conduct of trials in advanced cancer. These data suggest that interim and primary analyses of cancer trials can be conducted earlier in the course of the study, leading to a more rapid determination of futility or study benefit. In this way, ineffective therapies can be identified, permitting cross-over to better options, and effective therapies can be found more expeditiously. Early recognition translates directly to improved ethical standards and quality of life of enrolled subjects. In aggregate, more timely reporting of new, effective interventions accelerates the time to population benefit. Furthermore, because many of the inputs into clinical trial costing represent time-dependent and relatively fixed costs throughout the study, reduction in time-to-efficacy reporting in a clinical trial provides additional dramatic reductions in study costs [19].

The rapid expansion in the number of new therapeutic agents in oncology has created an urgent need for more efficient clinical trial processes. Traditional trial designs based on KM analysis are often slow and resource-heavy, making them inadequate for keeping pace with the rapid advancements in drug development. Adaptive trial designs, such as basket trials and umbrella trials, provide a partial solution by enabling the evaluation of multiple therapies or therapy combinations simultaneously, thus potentially reducing the time and cohort size needed to determine a drug's efficacy. Basket trials test the efficacy of one drug across different cancer types that share a common mutation [20], while umbrella trials test multiple drugs in a single cancer type but against different mutations [21]. In principle, CWTA could readily be incorporated into such studies to further reduce sample sizes and expedite an efficacy signal.

In any modeling exercise, some simplifications are required. For this study, all patients were accrued simultaneously, and the fully accrued trial was run from time 0, without simulating the staggered accrual which typically occurs in advanced cancer trials. However, sensitivity testing (data not shown) suggests that time to accrual does not meaningfully change the magnitude of these findings. In our simulations, we tested each month for statistical significance, and did not adjust for multiple testing; while there is a systematic bias toward underestimating time to positive study results, this bias was shared across each of the three analyses we conducted (CTWA, KM PFS, and KM OS), with the relative times to positive signals being the key output, rather than absolute times. In practice, such interim and final analyses would typically be pre-specified, in the case of KM PFS and KM OS by pre-specified events, or in the case of CWTA, specified by pre-defined number of health state transitions.

These data provide the conceptual framework for interrogating completed studies, in which datasets from advanced cancer trials could be studied to confirm the results of time to study positivity using actual trial data sequentially analyzed at similar intervals. Furthermore, in principle, the improvements in both sample size requirements and time to study efficacy signal should be evident in the setting of early stage cancer trials; we are pursuing additional simulation work to explore this possibility.

## 5. Conclusions

In aggregate, this study shows that by incorporating the entirety of the cancer trajectory including disease response, progression, and death, CWTA meaningfully reduces sample size requirements for clinical trials in the advanced cancer setting, particularly in those studies of less effective interventions. Furthermore, CWTA expedites the time-to-efficacy signals of cancer treatments compared to KM PFS and OS. Consequently, these data suggest that using CWTA rather than KM as the primary trial outcome has multiple efficiency advantages and the potential to reduce trial duration, cost, and numbers of patients needed to evaluate therapies in advanced cancer.

**Supplementary Materials:** The following supporting information are available on request from the corresponding author. Figure S1: Power as a function of sample size for Kaplan-Meier estimation of Progression-Free Survival and Overall Survival, and Chauhan Weighted Trajectory Analysis (CWTA) using a highly active control regimen; Figure S2: Time-to-efficacy signals comparison using a highly active control regimen; Table S1: Time-to-Efficacy Signals Comparison for (a) moderately active control regimen and (b) highly active control regimen; Table S2: 80% Power Sample Size Interpolation for Kaplan-Meier estimation of Progression-Free Survival and Overall Survival, and Chauhan Weighted Trajectory Analysis (CWTA) across a range of hazard ratios, using a highly active control regimen.

**Author Contributions:** Conceptualization, U.C. and J.R.M.; methodology, U.C. and J.R.M.; software, U.C. and D.M.; validation, U.C. and J.R.M.; formal analysis, U.C. and J.R.M.; investigation, U.C.; resources, U.C. and D.M.; data curation, U.C. and D.M.; writing—original draft preparation, U.C. and J.R.M.; writing—review and editing, D.M.; visualization, U.C. and J.R.M.; supervision, J.R.M. All authors have read and agreed to the published version of the manuscript.

**Funding:** This research received no external funding.

**Data Availability Statement:** The data that support the findings of the research are available on request from the corresponding author. The data are not publicly available due to privacy or ethical restrictions.

## 6. Supplementary Materials

**Table S1A.** Time-to-Efficacy Signals Comparison for a moderately active control regimen. This analysis applied a complete response rate ~5% and partial response rate ~30%. We performed one hundredfold / thousandfold simulations of time to study positivity as a function of sample size for Kaplan-Meier estimation of Progression Free Survival and Overall Survival, and for Chauhan Weighted Trajectory Analysis (CWTA) for a range of hazard ratios. CWTA demonstrated consistently shorter time to study positivity; this effect was most profound in studies with a highly effective intervention (that is, a low hazard ratio).

| HR | SS | Time-to-Efficacy (months; mean ± SD) | | | CWTA vs PFS | | CWTA vs OS | |
|---|---|---|---|---|---|---|---|---|
| | | CWTA | PFS | OS | %Δt | P-value | %Δt | P-value |
| 0.5 | 30 | 25.0 ± 14 | 34.8 ± 12 | 44.2 ± 8 | 28% | <0.0001 | 43% | <0.0001 |
| | 60 | 20.7 ± 13 | 29.1 ± 12 | 40.3 ± 8 | 29% | <0.0001 | 49% | <0.0001 |
| | 90 | 16.4 ± 12 | 26.8 ± 11 | 37.2 ± 8 | 39% | <0.0001 | 56% | <0.0001 |
| | 120 | 13.6 ± 9 | 23.6 ± 11 | 33.6 ± 7 | 43% | <0.0001 | 60% | <0.0001 |
| | 150 | 12.4 ± 9 | 22.6 ± 10 | 33.1 ± 7 | 45% | <0.0001 | 62% | <0.0001 |
| | 180 | 10.2 ± 8 | 21.1 ± 8 | 32.0 ± 6 | 52% | <0.0001 | 68% | <0.0001 |
| | 210 | 9.7 ± 7 | 19.6 ± 8 | 29.0 ± 5 | 51% | <0.0001 | 67% | <0.0001 |
| | 240 | 8.5 ± 7 | 20.4 ± 8 | 30.7 ± 6 | 58% | <0.0001 | 72% | <0.0001 |
| | 270 | 7.9 ± 6 | 18.6 ± 8 | 28.7 ± 5 | 58% | <0.0001 | 73% | <0.0001 |
| | 300 | 7.0 ± 5 | 17.0 ± 6 | 27.7 ± 5 | 59% | <0.0001 | 75% | <0.0001 |
| | 350 | 6.2 ± 5 | 15.9 ± 7 | 27.3 ± 5 | 61% | <0.0001 | 77% | <0.0001 |
| | 400 | 5.4 ± 4 | 14.9 ± 6 | 25.9 ± 5 | 64% | <0.0001 | 79% | <0.0001 |
| | 500 | 4.8 ± 4 | 13.6 ± 6 | 25.0 ± 4 | 65% | <0.0001 | 81% | <0.0001 |
| 0.6 | 30 | 28.7 ± 16 | 37.1 ± 12 | 45.8 ± 8 | 23% | <0.005 | 37% | <0.0001 |
| | 60 | 26.4 ± 15 | 32.8 ± 13 | 42.8 ± 9 | 20% | <0.005 | 38% | <0.0001 |
| | 90 | 20.4 ± 14 | 30.2 ± 13 | 39.4 ± 9 | 32% | <0.0001 | 48% | <0.0001 |
| | 120 | 19.1 ± 12 | 28.7 ± 12 | 38.1 ± 9 | 33% | <0.0001 | 50% | <0.0001 |
| | 150 | 19.7 ± 13 | 28.9 ± 13 | 35.7 ± 8 | 32% | <0.0001 | 45% | <0.0001 |
| | 180 | 16.8 ± 13 | 25.8 ± 12 | 32.9 ± 8 | 35% | <0.0001 | 49% | <0.0001 |
| | 210 | 14.2 ± 10 | 22.6 ± 12 | 32.2 ± 8 | 37% | <0.0001 | 56% | <0.0001 |
| | 240 | 12.5 ± 9 | 21.6 ± 11 | 31.6 ± 7 | 42% | <0.0001 | 61% | <0.0001 |
| | 270 | 11.6 ± 8 | 21.4 ± 9 | 31.5 ± 7 | 46% | <0.0001 | 63% | <0.0001 |
| | 300 | 10.2 ± 9 | 20.7 ± 10 | 29.9 ± 6 | 51% | <0.0001 | 66% | <0.0001 |
| | 350 | 8.9 ± 6 | 19.1 ± 9 | 29.2 ± 7 | 53% | <0.0001 | 70% | <0.0001 |
| | 400 | 8.9 ± 6 | 17.3 ± 8 | 27.7 ± 6 | 49% | <0.0001 | 68% | <0.0001 |
| | 500 | 6.2 ± 5 | 16.8 ± 8 | 27.3 ± 5 | 63% | <0.0001 | 77% | <0.0001 |
| 0.7 | 30 | 30.9 ± 16 | 37.6 ± 12 | 43.1 ± 6 | 18% | <0.05 | 28% | <0.0001 |
| | 60 | 26.1 ± 17 | 31.7 ± 14 | 41.7 ± 8 | 18% | <0.05 | 37% | <0.0001 |
| | 90 | 23.2 ± 16 | 30.8 ± 15 | 39.6 ± 10 | 25% | <0.005 | 41% | <0.0001 |
| | 120 | 23.5 ± 16 | 31.8 ± 14 | 39.7 ± 10 | 26% | <0.0005 | 41% | <0.0001 |
| | 150 | 22.6 ± 16 | 28.4 ± 14 | 35.6 ± 9 | 20% | <0.05 | 36% | <0.0001 |
| | 180 | 20.1 ± 15 | 26.7 ± 14 | 37.6 ± 10 | 25% | <0.005 | 46% | <0.0001 |
| | 210 | 19.3 ± 13 | 28.9 ± 14 | 37.4 ± 10 | 33% | <0.0001 | 48% | <0.0001 |
| | 240 | 19.0 ± 14 | 24.9 ± 12 | 35.7 ± 9 | 24% | <0.005 | 47% | <0.0001 |
| | 270 | 19.0 ± 13 | 24.6 ± 12 | 33.9 ± 9 | 23% | <0.005 | 44% | <0.0001 |
| | 300 | 16.1 ± 12 | 25.8 ± 13 | 34.3 ± 9 | 38% | <0.0001 | 53% | <0.0001 |
| | 350 | 14.8 ± 12 | 23.4 ± 12 | 32.8 ± 9 | 37% | <0.0001 | 55% | <0.0001 |
| | 400 | 13.7 ± 12 | 23.8 ± 12 | 33.3 ± 7 | 42% | <0.0001 | 59% | <0.0001 |
| | 500 | 12.2 ± 10 | 22.2 ± 9 | 31.1 ± 7 | 45% | <0.0001 | 61% | <0.0001 |
| | 30 | 27.4 ± 15 | 31.7 ± 12 | 43.6 ± 8 | 14% | 0.19 | 37% | <0.0001 |
| | 60 | 26.0 ± 16 | 30.0 ± 13 | 41.5 ± 9 | 14% | 0.18 | 37% | <0.0001 |
| | 90 | 24.3 ± 16 | 28.1 ± 13 | 40.1 ± 10 | 14% | 0.19 | 39% | <0.0001 |
| | 120 | 25.7 ± 17 | 29.4 ± 15 | 39.3 ± 10 | 13% | 0.22 | 35% | <0.0001 |
| | 150 | 24.1 ± 17 | 28.3 ± 15 | 38.2 ± 11 | 15% | 0.13 | 37% | <0.0001 |
| 0.8 | 180 | 24.5 ± 17 | 28.1 ± 15 | 37.8 ± 11 | 13% | 0.18 | 35% | <0.0001 |
| (1000- | 210 | 24.6 ± 17 | 28.5 ± 15 | 38.1 ± 11 | 14% | 0.14 | 36% | <0.0001 |
| fold) | 240 | 23.3 ± 16 | 28.2 ± 15 | 37.1 ± 11 | 17% | 0.05 | 37% | <0.0001 |
| | 270 | 22.9 ± 16 | 27.9 ± 15 | 36.7 ± 11 | 18% | <0.05 | 38% | <0.0001 |
| | 300 | 23.4 ± 16 | 27.2 ± 15 | 36.9 ± 11 | 14% | 0.12 | 37% | <0.0001 |
| | 350 | 21.9 ± 15 | 26.6 ± 14 | 36.3 ± 11 | 18% | <0.05 | 40% | <0.0001 |
| | 400 | 21.7 ± 15 | 26.7 ± 15 | 34.9 ± 10 | 19% | <0.05 | 38% | <0.0001 |
| | 500 | 20.5 ± 14 | 26.1 ± 14 | 35.1 ± 10 | 22% | <0.05 | 42% | <0.0001 |

**Table S1B.** Time-to-Efficacy Signals Comparison for highly active control regimen. This is a sensitivity analysis using a complete response rate ~10% and partial response rate ~50%. We performed one hundredfold / thousandfold simulations of time to study positivity as a function of sample size for Kaplan-Meier estimation of Progression Free Survival and Overall Survival, and for Chauhan Weighted Trajectory Analysis (CWTA) for a range of hazard ratios. CWTA demonstrated consistently shorter time to study positivity; this effect was most profound in studies with a highly effective intervention (that is, a low hazard ratio).

| HR | SS | Time-to-Efficacy (months; mean ± SD) | | | CWTA vs PFS | | CWTA vs OS | |
|---|---|---|---|---|---|---|---|---|
| | | CWTA | PFS | OS | %Δt | P-value | %Δt | P-value |
| | 30 | 21.4 ± 16 | 34.1 ± 12 | 45.0 ± 7 | 37% | <0.0001 | 52% | <0.0001 |
| | 60 | 14.7 ± 12 | 28.5 ± 12 | 38.7 ± 8 | 49% | <0.0001 | 62% | <0.0001 |
| | 90 | 11.4 ± 8 | 24.2 ± 11 | 34.8 ± 7 | 53% | <0.0001 | 67% | <0.0001 |
| | 120 | 8.3 ± 8 | 20.8 ± 8 | 34.4 ± 8 | 60% | <0.0001 | 76% | <0.0001 |
| | 150 | 7.9 ± 7 | 18.9 ± 7 | 30.2 ± 6 | 58% | <0.0001 | 74% | <0.0001 |
| 0.5 | 180 | 5.6 ± 4 | 18.1 ± 7 | 29.3 ± 5 | 69% | <0.0001 | 81% | <0.0001 |
| | 210 | 4.7 ± 4 | 16.3 ± 6 | 28.2 ± 5 | 71% | <0.0001 | 83% | <0.0001 |
| | 240 | 4.1 ± 3 | 15.2 ± 6 | 27.8 ± 5 | 73% | <0.0001 | 85% | <0.0001 |
| | 270 | 4.1 ± 3 | 14.7 ± 6 | 26.7 ± 5 | 72% | <0.0001 | 85% | <0.0001 |
| | 300 | 4.1 ± 4 | 14.8 ± 6 | 26.6 ± 5 | 72% | <0.0001 | 85% | <0.0001 |
| | 350 | 3.3 ± 3 | 13.5 ± 5 | 25.2 ± 5 | 76% | <0.0001 | 87% | <0.0001 |

| | 400 | 3.3 ± 2 | 12.3 ± 5 | 24.9 ± 5 | 73% | <0.0001 | 87% | <0.0001 |
|---|---|---|---|---|---|---|---|---|
| | 500 | 2.8 ± 2 | 11.9 ± 5 | 22.9 ± 4 | 76% | <0.0001 | 88% | <0.0001 |
| 0.6 | 30 | 24.4 ± 17 | 32.5 ± 12 | 45.5 ± 8 | 25% | <0.005 | 46% | <0.0001 |
| | 60 | 18.4 ± 15 | 28.8 ± 14 | 41.4 ± 9 | 36% | <0.0001 | 56% | <0.0001 |
| | 90 | 19.3 ± 14 | 26.4 ± 12 | 38.4 ± 8 | 27% | <0.0005 | 50% | <0.0001 |
| | 120 | 13.8 ± 11 | 25.7 ± 11 | 35.4 ± 8 | 46% | <0.0001 | 61% | <0.0001 |
| | 150 | 12.9 ± 10 | 24.2 ± 11 | 34.2 ± 8 | 47% | <0.0001 | 62% | <0.0001 |
| | 180 | 8.8 ± 7 | 19.5 ± 9 | 32.2 ± 7 | 55% | <0.0001 | 73% | <0.0001 |
| | 210 | 11.0 ± 8 | 19.6 ± 9 | 30.9 ± 7 | 44% | <0.0001 | 64% | <0.0001 |
| | 240 | 8.0 ± 7 | 18.0 ± 8 | 29.3 ± 7 | 55% | <0.0001 | 73% | <0.0001 |
| | 270 | 8.3 ± 7 | 19.1 ± 9 | 29.7 ± 7 | 57% | <0.0001 | 72% | <0.0001 |
| | 300 | 6.6 ± 6 | 16.9 ± 7 | 27.6 ± 6 | 61% | <0.0001 | 76% | <0.0001 |
| | 350 | 5.5 ± 5 | 16.0 ± 7 | 27.0 ± 6 | 66% | <0.0001 | 80% | <0.0001 |
| | 400 | 5.8 ± 5 | 15.1 ± 7 | 26.4 ± 6 | 61% | <0.0001 | 78% | <0.0001 |
| | 500 | 4.6 ± 3 | 13.9 ± 6 | 25.7 ± 5 | 67% | <0.0001 | 82% | <0.0001 |
| 0.7 | 30 | 24.6 ± 15 | 34.8 ± 12 | 44.3 ± 8 | 29% | <0.0005 | 44% | <0.0001 |
| | 60 | 22.3 ± 15 | 31.5 ± 15 | 41.5 ± 10 | 29% | <0.0005 | 46% | <0.0001 |
| | 90 | 20.4 ± 17 | 30.4 ± 16 | 41.0 ± 10 | 33% | <0.0001 | 50% | <0.0001 |
| | 120 | 18.9 ± 15 | 28.3 ± 13 | 38.7 ± 9 | 33% | <0.0001 | 51% | <0.0001 |
| | 150 | 19.6 ± 16 | 26.4 ± 13 | 38.2 ± 10 | 26% | <0.005 | 49% | <0.0001 |
| | 180 | 17.6 ± 14 | 28.7 ± 13 | 34.9 ± 10 | 39% | <0.0001 | 50% | <0.0001 |
| | 210 | 16.0 ± 13 | 26.3 ± 13 | 34.7 ± 9 | 39% | <0.0001 | 54% | <0.0001 |
| | 240 | 14.8 ± 12 | 22.8 ± 13 | 33.3 ± 9 | 35% | <0.0001 | 55% | <0.0001 |
| | 270 | 14.2 ± 11 | 22.2 ± 11 | 33.2 ± 9 | 36% | <0.0001 | 57% | <0.0001 |
| | 300 | 13.1 ± 9 | 22.2 ± 10 | 31.6 ± 9 | 41% | <0.0001 | 58% | <0.0001 |
| | 350 | 12.5 ± 10 | 21.0 ± 10 | 31.6 ± 8 | 41% | <0.0001 | 61% | <0.0001 |
| | 400 | 10.3 ± 10 | 19.1 ± 11 | 31.2 ± 9 | 46% | <0.0001 | 67% | <0.0001 |
| | 500 | 9.7 ± 7 | 17.3 ± 9 | 27.3 ± 7 | 44% | <0.0001 | 64% | <0.0001 |
| 0.8 (1000-fold) | 30 | 23.1 ± 17 | 30.8 ± 13 | 43.3 ± 9 | 25% | <0.05 | 47% | <0.0001 |
| | 60 | 22.8 ± 17 | 28.7 ± 14 | 40.3 ± 10 | 21% | 0.06 | 44% | <0.0001 |
| | 90 | 22.6 ± 17 | 28.6 ± 14 | 40.5 ± 10 | 21% | <0.05 | 44% | <0.0001 |
| | 120 | 21.4 ± 17 | 28.9 ± 15 | 38.8 ± 11 | 26% | <0.05 | 45% | <0.0001 |
| | 150 | 22.2 ± 16 | 27.8 ± 15 | 38.2 ± 11 | 20% | <0.05 | 42% | <0.0001 |
| | 180 | 22.5 ± 17 | 27.9 ± 14 | 37.9 ± 11 | 19% | <0.05 | 41% | <0.0001 |
| | 210 | 22.4 ± 17 | 27.1 ± 15 | 37.9 ± 11 | 17% | 0.05 | 41% | <0.0001 |
| | 240 | 21.4 ± 16 | 26.4 ± 15 | 37.0 ± 11 | 19% | <0.05 | 42% | <0.0001 |
| | 270 | 21.8 ± 16 | 27.6 ± 15 | 36.8 ± 11 | 21% | <0.05 | 41% | <0.0001 |
| | 300 | 20.2 ± 15 | 26.8 ± 14 | 35.9 ± 11 | 25% | <0.005 | 44% | <0.0001 |
| | 350 | 19.2 ± 15 | 25.9 ± 14 | 35.3 ± 11 | 26% | <0.005 | 46% | <0.0001 |
| | 400 | 18.3 ± 14 | 25.2 ± 14 | 34.3 ± 11 | 27% | <0.005 | 47% | <0.0001 |
| | 500 | 16.2 ± 13 | 24.3 ± 13 | 32.9 ± 10 | 33% | <0.0001 | 51% | <0.0001 |

**Table S2.** 80% Power Sample Size Interpolation for Kaplan-Meier estimation of Progression-Free Survival and Overall Survival, and Chauhan Weighted Trajectory Analysis (CWTA) across a range of hazard ratios, using a highly active control regimen. This is a sensitivity analysis using a complete response rate ~10% and partial response rate ~50%. CWTA reduces the required sample size for 80% power by 9% to 22% versus KM PFS and 10% to 14% versus KM OS.

| HR | Sample Size required for 80% Power | | | Sample Size Reduction using CWTA | |
|---|---|---|---|---|---|
| | CWTA | PFS | OS | vs PFS | vs OS |
| 0.5 | 50 | 55 | 58 | 9% | 14% |
| 0.6 | 65 | 74 | 72 | 12% | 10% |
| 0.7 | 114 | 136 | 128 | 16% | 11% |
| 0.8 | 270 | 344 | 305 | 22% | 11% |

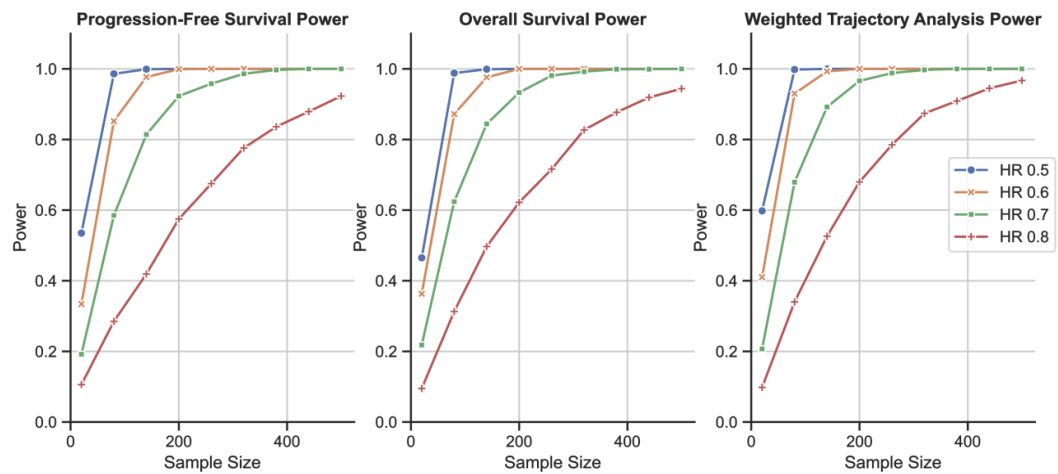

**Figure S1.** Power as a function of sample size for Kaplan-Meier estimation of Progression-Free Survival and Overall Survival, and Chauhan Weighted Trajectory Analysis (CWTA) using a highly active control regimen. This is a sensitivity analysis using a complete response rate ~10% and partial response rate ~50%. We performed thousandfold simulations of power as a function of sample size for Kaplan-Meier estimation of Progression-Free Survival and Overall Survival, and Chauhan Weighted Trajectory Analysis (CWTA) across a range of hazard ratios. CWTA demonstrated consistently higher power, reflecting a smaller sample size requirement during trial design. Assuming a trial designed to achieve a power of 0.8, WTA reduced sample size requirements by 9% to 22% when compared to KM PFS, and 11% to 13% when compared to KM OS; the sample size reductions were most marked for studies designed for interventions with higher hazard ratios (that is, less effective interventions).

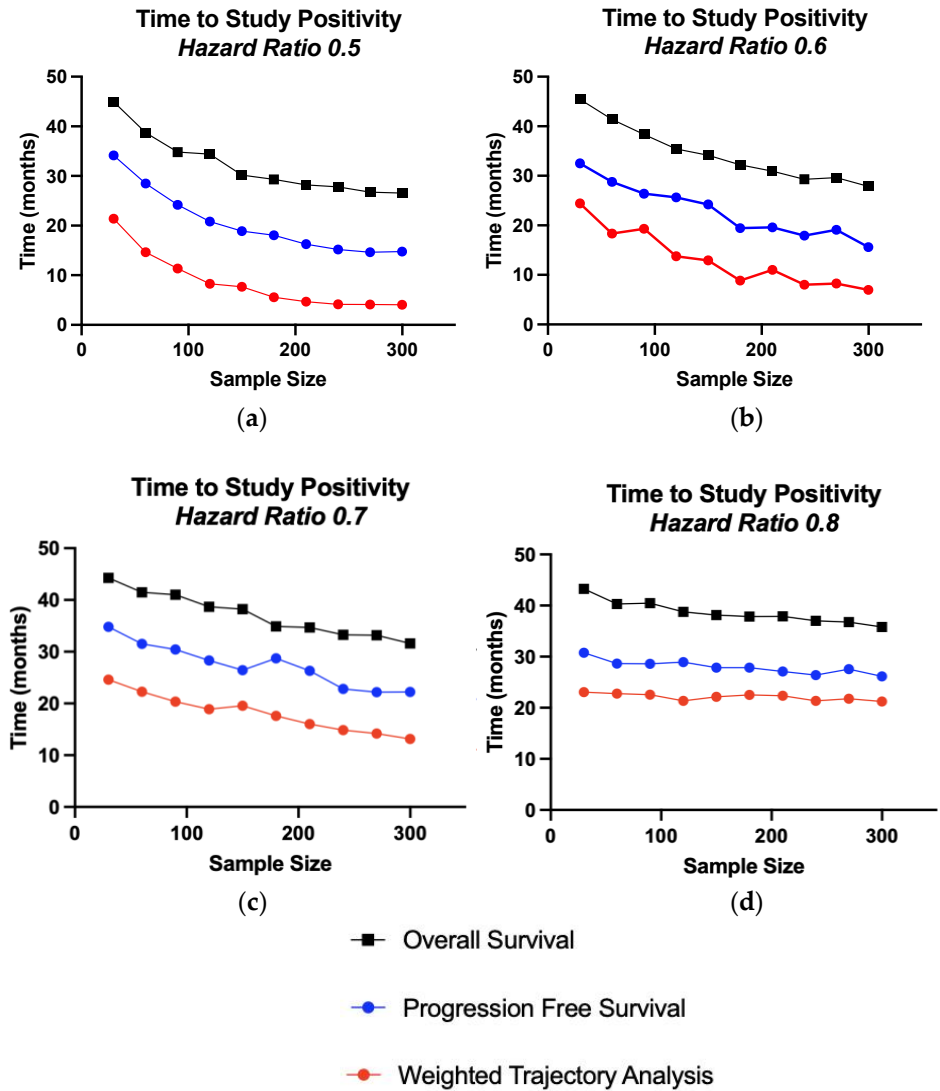

**Figure S2.** Time-to-efficacy signals comparison using a highly active control regimen. This is a sensitivity analysis using a complete response rate ~10% and partial response rate ~50%. One hundredfold / thousandfold simulations of time to study positivity as a function of sample size for Kaplan-Meier estimation of Progression Free Survival and Overall Survival, and for Chauhan Weighted Trajectory Analysis (CWTA) for a range of hazard ratios. CWTA demonstrated consistently shorter time to study positivity; this effect was most profound in studies with a highly effective intervention (that is, a low hazard ratio).